\documentstyle[aps,prl]{revtex}
\def\beq{\begin{equation}}
\def\eeq{\end{equation}}
\begin{document}                
\title{Parity Effect and Charge Binding Transition in Submicron
       Josephson Junction Arrays}
\author{M. V. Feigel'man, S. E. Korshunov and A. B. Pugachev}
\address{L. D. Landau Institute for Theoretical Physics, Moscow 117940, RUSSIA}
\date{\today}
\maketitle

\begin{abstract}
We  reconsider the issue of Berezinskii-Kosterlitz-Thouless (BKT) transition
into an insulating state in the Coulomb-dominated Josephson junction arrays.
We show that previously predicted picture of the Cooper-pair BKT transtion
at $T = T_2$ is valid only under the condition that $T_2$ is considerably
below the parity-effect temperature $T^* \approx 0.1 \Delta$ and
even in this case it is not a rigorous phase transition but only a
crossover, whereas the real phase transition takes place at $T_1\approx T_2/4$.
Our theory is in agreement with available experimental data on Coulomb-dominated
Josephson arrays and also sheds some light
on the origin of unusual reentrant temperature dependence of resistivity
in the array with nearly-criticial ratio  $E_C/E_J$.

\end{abstract}

{\bf 1.} Two-dimensional arrays of micron-scale superconducting islands are extensively
studied during last years, both experimentally
\cite{delft1,delft2,mit,chalmers}
 and theoretically \cite{FS,duality,mesobook94}. It is well-established now that their
low-temperature behaviour is
determined by the competition between
Josephson coupling energy $E_J$ and effective charging energy $E_C = e^2/2C$,
where $C$ is some relevant electric capacitance (to be discussed below).
Macroscopic superconductive coherence was observed at low tempratures
 in the arrays with  $E_J \gg E_C$, whereas arrays with $E_J \ll E_C$
show insulating behaviour at $T \rightarrow 0$. At nearly-critical value
of the ratio $x = E_J/E_C \sim x_{cr}$  direct transition between
superconductive (SC) and insulating (I) behavior as function of $x$ was observed
 in zero magnetic field \cite{delft2}.
MOrevoer, very weak magnetic field $B \leq 1 G$ (producing small fractions of flux
 quantum per unit cell of the array) was shown to switch arrays with
$x \approx x_{cr}$ between SC-like and I -like behavior as function of
temperature; recently very interesting intermediate
region was found \cite{delft2} where resistance $R(T)$ is basically constant in
the temperature range $10 mK \leq T \leq 200 mK$, which indicates the existence
of a "2D metal" state sandwiched between SC and I phases.

The above-mentioned
 basic properties  of 2D arrays are in qualitative agreement with available
theories \cite{FS,mesobook94}
 (except for the recently observed 2D metal state);
however several important features are not understood yet. In particular,
resistance of the insulating arrays shows purely activated behaviour
$R(T) \propto \exp(E_a/T)$ with constant activation energy $E_a$ through
the whole temperature interval studied \cite{delft2,mit,chalmers}, whereas
theoretically the charge binding Berezinskii-Kosterlitz-Thouless
(BKT) transition
\cite{BKT,K}
from the conducting to insulating phase is expected
to occur \cite{FS} at the temperature $T_2 \approx E_C/\pi$.
Such a transition should occur due to nearly-logarithmic form
of  Coulomb  interaction between charges
in the arrays with self-capacitance of islands $C_0$ very small compared to
the inter-island (junction)  capacitance $C$.
In the currently studied arrays the ratio $C/C_0 \sim 100$
(as measured at very low temperatures, about $ 10 mK$, cf. e.g. \cite{delft2}),
which should result in the logarithmic interaction throughout the whole
array (effecitve length of interaction $\Lambda$ should be estimated with the
account of 3D nature of electric field, which leads \cite{delft2,delft0} to
$\Lambda \sim C/C_0 \sim 100$) and, consequently, to the charge binding BKT
transition.

In the case of islands in SC state and under the condition $E_J \ll E_C$
the temperature of this transtion was estimated \cite{FS} as
$T_2 \approx E_C/\pi$, whereas in the case of normal islands (i.e. with
superconductivity suppressed by magnetic field) it is expected to be 4 times
lower: $T_1 \approx E_C/4\pi$ due to the twofold decrease of an elementary
charge available. Nevertheless
no indication of such a transtion in array of SC islands
 was found experimentally (except in very recent
preprint \cite{japan}, which is discussed below).
Another surprising feature observed in \cite{delft2} is nonmonotoneous
("reentrant") temperature behaviour of resistance $R(T)$ of the
array with a nearly-critical $E_J/E_C$ ratio at $T \leq 200 mK$.

In the present Letter we show that the above experimental observations
can be naturally understood once the temperature dependence of effective
Coulomb interation between the charges in the array is taken into account
properly. In a very broad sense our analysis follows the ideas of Efetov
\cite{efetov} who established the background for the description of quantum
fluctuations in granular superconductors;
namely, we consider screening of Cooper-pair
Coulomb interaction by normal quasiparticles existing in each superconductive
island at finite temperatures.
However we believe that Efetov's
treatment of the effect he proposed  technically  was not quite correct,
thus we present here another theoretical approach to the same problem.

Our main qualitative result can be formulated as follows: at the temperatures
above
the so-called parity-effect \cite{parity1,parity,nazarov,glazman} temperature
$T^* \approx\Delta/\ln{\cal M}\ll\Delta$
[where ${\cal M} = V\nu(0)\sqrt{8\pi T\Delta} \sim 10^4-10^5$, $V$ is the volume of
the island and $\nu(0)$ is the density of states at the Fermi level in absense
of superconductivity]
the presence of thermal
quasiparticles [with the number $\sim {\cal M}\exp(-\Delta/T) \gg 1$]
 on each island
 excludes any possibility to observe Cooper-pairs
 BKT transtion at $T_2$. Since in the most of arrays studied till now
the above-defined characterisitic temperature $T_2$ was in the range
0.3-0.5 K, whereas parity effect temperature $T^* \approx 0.2 K < T_2$,
the absense of anything like BKT transtion near $T_2$ is quite natural
(measurements below $T^*$ were not possible in these arrays
\cite{mit,chalmers}
 since $R(T)$ becomes immeasurably high [$\geq 10^9 Ohm$]).
On the other hand single-electron BKT transtion is a completely
different issue: we do expect such a transition to be observable
at approximately the same temperature $ T_1 \approx T_{2}/4$ as
in the arrays with islands in the normal state.

{\bf 2.} We proceed now to the derivation of our results, and will follow, with one
important modification, Ref. \cite{FS}.
In the limit when charge tunneling is weak and is important only
for the establishment of thermodynamic equilibrium
an array of superconducting islands can be described by a classical partition
function of a form:
\begin{equation}
Z=\sum_{\{n\}}^{}
\exp\left[-\frac{1}{2}\sum_{i,j} G_{ij}n_i n_j
-\frac{D}{T}\sum_{j}\frac{1-(-1)^{n_j}}{2}\right];~~~~
G_{ij}=\frac{e^2}{T}C^{-1}_{ij}                                 \label{S1}
\end{equation}
The first term in the exponent in Eq. (\ref{S1}) stands for the electrostatic
energy of the array
which in the case when only mutual capacitance of nearest islands $C$ is
of importance corresponds to the logarithmic interaction of the charges in
two-dimensional array:
\begin{equation}
G_{i=j}-G_{ij}\approx \frac{2 E_C}{T}\left(\frac{1}{2\pi}\ln R_{ij}
+\frac{1}{4}\right);~~~~E_C=\frac{e^2}{2C}
                                                               \label{S2 }
\end{equation}
whereas the second term describes the dependence of a free energy
of a superconducting island on the parity of the number of electrons $n_j$ on
this island \cite{parity1,nazarov,glazman}.

The free energy difference $D(T)$
between the islands with odd and even number of electrons can be expressed as
\begin{equation}
D(T)=-T\ln\tanh(\Omega_{oe}/{T})                                   \label{S3}
\end{equation}
where
$\Omega_{oe}=-T\ln(Z_{odd}/Z)$ and $Z_{odd}$ is the "odd grand canonical
partition function" introduced in Ref.\cite{parity1}
for the study of parity effect.
Accurate expression for the function $\Omega_{oe}(T)$ can be found in
\cite{nazarov}, but for $T\ll\Delta$ a good approximation is
given by
\begin{equation}
\Omega_{oe}(T)/{T}\approx {\cal M}e^{-\Delta/T}; \quad
{\cal M} = V\nu(0)\sqrt{8\pi T\Delta}
\label{S4} \end{equation}
The ratio $\Omega_{oe}/T$ in that limit is proportional to
the number of thermally exited quasiparticles on one island.

In terms of statistical mechanics partition function (\ref{S1})
defines a lattice Coulomb gas in which the fugacities of odd charges
$Y_{}=\exp(-D/T)$ differ from the fugacities of even charges
(which are equal to one).
Comparison of Eq. (\ref{S4}) with Eq. (\ref{S3})
shows then that for $T\ll T^*=\Delta/\ln{\cal M}$
the parity-dependent free energy difference
$D(T)\approx \Delta-T\ln{\cal M}\gg T$ and $Y_{}\ll1$,
whereas in the opposite limit $T\gg T^*$ the quantity
$D(T)$ becomes exponentially small and $Y_{}$ is very close to one.

Previously it has been assumed \cite{FS} that in the regime when island
charges behave as classical variables the main difference between the array of
normal islands and the array of superconducting islands is that in the array
of
normal islands the charge of each island is quantized in units of $e$, whereas
in the array of superconducting islands the charge is quantized in units of
$2e$. The consequence for the array the electrostatic properties of which are
dominated by mutual capacitance of nearest neighbours is that the temperature
$T_2$ of the BKT transition in the array of
superconducting islands (appearance of free  double charges) should be exactly
four times higher than the temperature
$T_1\sim E_C/4\pi$ of the BKT transition in the
analogous array of normal islands (appearance of free single charges).
Comparison with Eq. (\ref{S1}) shows
that such description of the array of superconducting islands
would be correct only in the limit of $D(T)/T\rightarrow\infty$.
Since $D(T)$ is always finite this description turns out to be misleading.
The behaviour of the array at temperatures close to $T_2 = E_C/\pi$
depends qualitatively on the relation between $T_2$ and $T^*$; we consider
both cases in turn.

{\bf 3.} At $T_2\geq T^*$ an array of the superconducting islands is
described by practically the same partition function  as an array of the
normal islands, since $D(T_2) << T_2$ in that case.
 The phase transition into insulating state in such system can be associated
 with the binding of the charges $\pm 1$ into neutral
pairs; it takes place
at the temperature $T_1$ which is slightly lower than the simple estimate
$T_1^{(0)}=E_C/4\pi$ which can be obtained by comparison of the single charge
energy with its entropy.
The difference between $T_1$ and $T_1^{(0)}$ is
related to the renormalization of charge interaction by bound pairs of charges
and decreases with decrease in fugacities.
The appearance of the free single charges
induces the screening of the Coulomb interaction for all types of charges and
therefore the double charges also are free at $T > T_1$.
Not even a trace of a separate phase transition related to debounding of
double charges can be expected to be observed in such a situation,
which was realized in the experiments  \cite{delft1,delft2,mit,chalmers}.

{\bf 4.} In the opposite case $T_2 < T^*$ there is a range of temperatures
$T_2 < T < T^*$ where
fugacity  of single charges $Y$ is much smaller than one. This leads to the
increase of phase transition temperature,  but it still has to  remain smaller
than $T_1^{(0)}$.
The  difference  with  the  case  of  $Y\approx  1$  is  that for $Y\ll 1$ the
concentration of free single charges $n_1$  remains small even
at the temperatures considerably higher than $T_1$.
In the region $T_1 < T < T_2$ it can be estimated with the use of
the standard Debye-H\"{u}ckel approximation
which gives for $n_1$ the self-consistent equation:
\begin{equation}
n_1=2Y\exp\left\{-\frac{1}{2}\int_{}^{}\frac{d^2 {\bf q}}{(2\pi)^2}
\frac{1}{K[2(1-\cos q_x)+2(1-\cos q_y)]+n_1}\right\};
                                                                 \label{SCA}
\end{equation}
(where $K={T}/{2E_C}$)
the solution of which for small $n_1$ can be expressed as
\begin{equation}
n_1 \approx 2\exp\left\{-\frac{D(T)+
aE_C - [\ln(4E_C/T)]E_C/4\pi}{T-T_1^{(0)}}\right\}
                                                                 \label{S5}
\end{equation}
where $a = 0.276...$.
The main effect of Coulomb interaction is seen in that it produces
singularity at $T = T_1^{(0)}$ in the exponent in Eq.(\ref{S5}).
If the shift of the phase transition temperature is taken
into acccount $T_1^{(0)}$ should be  substituted by $T_1$.
Note that second and third terms in the
numenator of the exponent almost canceal each other
in the relevant range of parameters.

For $D(T)\gg T$ the screening
of  the  interaction  is noticable  only  on  the  large scales and the
concentration of  free double  charges  also remains  small.  On the other
hand at the temperature $T_2=4T_1\approx E_C/\pi$ the free double charges have
to appear even when $Y=0$ due to mutual influence of pairs of double charges
(cf. with Ref.\cite{FS}).
That means  that  for  $D(T)\gg  T$  in  the  vicinity  of $T_2$ there
occurs a crossover characterized by a prolifiration of free double charges.

Close to the transition temperature $T_1$ the self-consistent approximation
is no longer valid and more advanced methods should be used.
It is easy to show that when only the single
and double charges are taken into account,
the Coulomb gas model described by partition function (\ref{S1}) becomes
isomorphic (in continuous approximation) to the sine-Gordon model defined
by the Hamiltonian:
\begin{equation}
H=\int d^2{\bf r}\left[\frac{K}{2}(\nabla \theta)^2-2Y\cos
\theta-2\cos2\theta\right];~~K=\frac{T}{2E_C}                    \label{S6}
\end{equation}
The renormalization group equations for the Hamiltonian (\ref{S6})
can be found in Ref.\cite{RG}. As can be expected their solution shows that
for temperatures lower than $T_2$ the presence of double charges (in form of
neutral pairs) does not introduce any qualitative changes.
In the close vicinity of $T_1$ the temperature dependence of $n_1$
deviates from the self-consistent result (\ref{S5}) and follows the
 standard BKT critical behaviour
\cite{K} with
\beq
n_1(T) \propto \exp{\left[-\frac{b}{\sqrt{1-T/T_1}}\right]}
\label{BKT}
\eeq
where $b$ is of the order of unity.
The array's linear dc resistance should be inversely proportional to
the density of free charges $n_1$. Taking into account Eqs.(\ref{S5},
\ref{BKT}), we get an estimate for this resistance at the temperatures
near $T_1$:
\beq
\ln \frac{R(T)}{R_1}  \approx  \min\left[
\frac{D(T)}{T-T_1}, \frac{b}{\sqrt{4\pi}}\left(\frac{E_C}{T-T_1}\right)^{1/2} \right]
\label{rho}
\eeq
where $R_1$ is inversely proportional to the probability of tunnelling
event which is only weakly dependent on the temperature:
$R_1 \sim R_n{\cal M}$,
where $R_n$ is the normal-state tunnelling resistance (cf. \cite{ambeg}).

{\bf 5.}
The representation (\ref{S6}) is also useful for comparison between our
results and Efetov's treatment of screening by quasiparticles \cite{efetov},
 which can be expressed just as the replacement of the original capacitance
matrix $C_{ij}$ by the "effective" one defined (at $T\ll \Delta$) as
\beq
C^{eff}_{ij} =  C_{ij} +
\delta_{ij}\cdot V \nu(0)(2e)^2 \sqrt{\frac{2\pi\Delta}{T}}
e^{-\Delta/T}
\label{Ceff}
\eeq
Let us now formally expand the second term in Eq.(\ref{S6}) up to the
 second order
in $\theta$, and neglect the rest of terms.
One can easily see that the expression obtained in this way
for the effective interaction between 2e charges (generated by expansion
of the partition function in powers of the $\cos2\theta$ term)
 would coinside with the one obtained by inversion of
 the effective capacitance (\ref{Ceff}).
Physically the above formal operation would mean neglect of the discrete
nature of electric charge, which could be reasonable
if  charge transport between islands could proceed via some "classical"
channels able to provide charge in continuous amounts (like
 charging of macroscopic electric capacitor by external voltage source).
It is not the case for submicron arrays with tunnel junctions where
charges of islands can change only by $1e$ quantum (which is just reflected
by periodic nature of the second term in Eq.(\ref{S6})).  However, basic
 qualitative feature of Efetov's result - screening of Cooper pair
charges by normal exitations - remanins valid, in spite of the absense of
any simple notion like an effective capacitance matrix.

{\bf 6.}
It follows from our results, that in order to observe some growth of
effective activation energy [defined as $E_a= d\ln R(T)/d(1/T)$] with
$T$ approaching $T_2$ from above, one needs to use array with the
Coulomb energy $E_C$ few times below than $\Delta$, so that $T_2 \leq T^*$.
This conclusion is in complete agreement with recent experimental data
\cite{japan}, where some moderate growth of $E_a(T)$ in the temperature
range around 0.2-0.3 K was observed. This experiment  differs
from few previous ones of the same type \cite{delft2,mit,chalmers} (where
constant $E_a$ was observed) by lower values of  $E_C$ and a bit higher
reported $\Delta$. On the other hand, we do not agree with the
interpretation of that $E_a(T)$ growth as precursor of the BKT transtion
at $T_2$, given in \cite{japan} for their SC arrays. As follows from
our results, no such a transtion exists at $T_2$,  which agrees with a
rather modest (compared to BKT behaviour) growth of $E_a(T)$ observed in
\cite{japan} at $T \sim T_2$. Note that
the agreeement between the data reported in \cite{japan} for normal arrays
and the expected genuine BKT transition at $T_1$ is much better than
the above-mentioned comparison for SC arrays.

The above theoretical results point out that any analysis of
experimental data on superconductor-insulator transtion
in artificial arrays of superconducting islands (as well as in the
 dirty thin films near SC-I transtion)
should take into account an existence of a characteristic temperature scale
of the parity effect $T^*$
(note that $T^*$ is magnetic-field dependent and strongly suppressed by
the fields of the order of $H_{c2}$).
In particular, the behavior of I-V characteristics
in the intermediate temperature range $T^* < T < \Delta$
cannot be unambigously related to the genuine ground-state properties
of the system, as is examplified by non-monotonous $R(T)$ behaviour observed
in \cite{delft2} for a nearly critical ratio $E_J/E_C$. We interpret this
unusual behaviour as follows: at moderately low $T$
screening by quasiparticles is
effective and reduces Coulomb repulsion of Cooper pairs, leading to
the decrease of $R(T)$ behaviour; at still lower  $T$ this screening
is gone, Coulomb repulsion increases and effective ratio $E_J/E^{eff}_C$
enters the "insulative" part of the phase diagram, leading to the
increase of $R(T)$ at further $T$ decrease.

We are grateful to  V. B. Geshkenbein, A. Kitaev, N. B. Kopnin,
A. I. Larkin, J. E. Mooji, Yu. V. Nazarov, G. Sch\"{o}n, M.Skvortsov and
H. van der Zant for many helpful discussions. Financial support from  INTAS-RFBR grant
\# 95-0302 (M.V.F.)
and Swiss National Science Foundation collaboration grant \# 7SUP J048531
(M.V.F. and A.V.P.),
as well as the DGA grant \# 94-1189 (M.V.F.)
and RFBR grant \# 96-02-18985 (S.E.K) is gratefully acknowledged.

\end{document}